\pdfoutput=1

\documentclass[11pt]{article}

\usepackage{EMNLP2023}

\usepackage{times}
\usepackage{latexsym}
\usepackage{float}
\usepackage{multirow}
\usepackage{color, colortbl}
\usepackage{arydshln}
\usepackage{array}
\usepackage{booktabs}
\usepackage{graphicx}
\usepackage{caption}
\usepackage{subcaption}
\usepackage{bbding}

\usepackage[T1]{fontenc}

\usepackage[utf8]{inputenc}

\usepackage{microtype}

\usepackage{inconsolata}

\definecolor{Gray}{gray}{0.9}
\newcolumntype{g}{>{\columncolor{Gray}}c}
\newcommand{\newsreclib}{\texttt{NewsRecLib}}
\newcommand{\rparagraph}[1]{\vspace{1.4mm}\noindent\textbf{#1}}

\title{NewsRecLib:\\ A PyTorch-Lightning Library for Neural News Recommendation}

\author{Andreea Iana\textsuperscript{1}, Goran Glava{\u s}\textsuperscript{2}, Heiko Paulheim\textsuperscript{1} \\ 
       \textsuperscript{1} Data and Web Science Group, University of Mannheim, Germany \\ 
        \textsuperscript{2} Center For Artificial Intelligence and Data Science, University of Würzburg, Germany \\
        \texttt{\{andreea.iana, heiko.paulheim\}@uni-mannheim.de} \\
        \texttt{goran.glavas@uni-wuerzburg.de}
        }

\begin{document}
\maketitle

\begin{abstract}
\newsreclib{}\footnote{\url{https://github.com/andreeaiana/newsreclib}} is an open-source library based on Pytorch-Lightning and Hydra developed for training and evaluating neural news recommendation models. The foremost goals of \newsreclib{} are to promote \textit{reproducible research} and \textit{rigorous experimental evaluation} by (i) providing a unified and highly configurable framework for exhaustive experimental studies and (ii) enabling a thorough analysis of the performance contribution of different model architecture components and training regimes. \newsreclib{} is highly modular, allows specifying experiments in a single configuration file, and includes extensive logging facilities. Moreover, \newsreclib{} provides out-of-the-box implementations of several prominent neural models, training methods, standard evaluation benchmarks, and evaluation metrics for news recommendation. 
\end{abstract}

\section{Introduction}

Personalized news recommendation
has become ubiquitous for customizing suggestions to users' interests \cite{li2019survey, wu2023personalized}. 
In recent years, there has been a surge of effort towards neural content-based recommenders. With increasingly complex neural architectures able to ever more precisely capture users' content-based preferences, neural recommenders quickly replaced traditional recommendation models as the go-to paradigm for news recommendation. 

Despite the abundance of model designs, research on neural news recommenders (NNRs) suffers from two major shortcomings: (i) a surprising amount of non-reproducible research \cite{ferrari2021troubling} and (ii) unfair model comparisons \cite{ferrari2019we,sun2020we}. The former is, on the one hand, due to many NNR implementations not being publicly released  \cite{sertkan2022diversifying}. Existing open source repositories, on the other hand, expose a multitude of programming languages, libraries, and implementation differences, hindering reproducibility and extensibility \cite{said2014comparative}.
Moreover, a lack of transparency in terms of evaluation datasets, experimental setup and hyperparameter settings, as well as the adoption of ad-hoc evaluation protocols, further severely impede direct model comparisons. Many personalized news recommenders have been evaluated on proprietary datasets (e.g., Bing News \cite{wang2018dkn}, MSN News \cite{wu2019naml,wu2019nrms}, News App \cite{qi2022news}). Even the models trained on the more recently introduced open benchmarks (e.g., Adressa \cite{gulla2017adressa}, MIND \cite{wu2020mind}) cannot be directly compared due to the lack of standard dataset splits and evaluation protocols \cite{wu2021empowering,zhang2021amm,gong2022positive,wang2022news}. Even more concerning, crucial details regarding the setup of the experiments are regularly omitted from the publications or hard-coded without explanation.

It is thus particularly difficult to evaluate the impact of specific components in NNR architecture and training (e.g., news encoder, user modeling, training objectives) on the overall performance of the model \cite{iana2023simplifying}. Many models simultaneously change multiple components in both the news and the user encoder, while carrying out only partial ablation studies or evaluating against suboptimal baselines \cite{rendle2019difficulty}.

In this work, we introduce \newsreclib{}, an open source library for NNRs, to remedy these critical limitations.\footnote{The library is licensed under a MIT license.} 
\newsreclib{} aims to facilitate reproducible research and comprehensive experimental studies, using an end-to-end pipeline powered by a single configuration file that specifies a complete experiment -- from dataset selection and pre-processing over model architecture and training to evaluation protocol and metrics. \newsreclib{} is built based on the following guiding principles:

\begin{figure*}[t]
  \centering
  \includegraphics[width=0.9\textwidth]{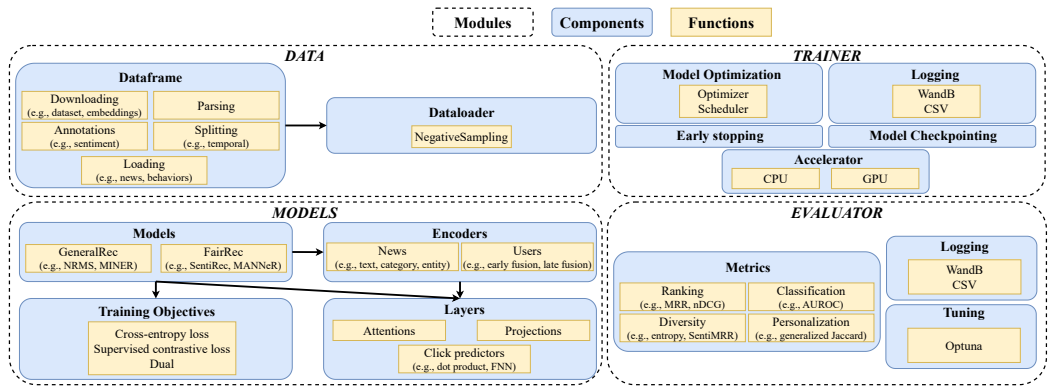}
  \caption{Illustration of the \newsreclib{} framework.}
  \label{fig:framework}
\end{figure*}

\rparagraph{Modularity and extensibility.}
With PyTorch Lightning \cite{Falcon_PyTorch_Lightning_2019} as its backbone, \newsreclib{} is designed in a modular fashion, with core individual components being decoupled from one another. This enables mixing and matching different modules, as well as seamlessly integrating new ones.

\rparagraph{Easy configurability and reproducibility.}
\newsreclib{} is powered by Hydra \cite{Yadan2019Hydra}, in which each experiment is defined through a single configuration file 
composed from the configurations of specific pipeline components. The configuration of every experiment is automatically stored at the start of the run and as such trivially enables reproducibility.

\rparagraph{Logging and profiling.}
The library supports multiple standard tools (e.g., WandB \cite{wandb}, Tensorboard \cite{abadi2016tensorflow}) for extensive logging, monitoring, and profiling of experiments with neural models -- in terms of losses, evaluation metrics, runtime, memory usage, and model size.

\rparagraph{}
Overall, \newsreclib{} is designed to support the development and benchmarking of NNRs as well as the specific analysis of contributions of common components of the neural recommendation pipelines. In this paper, we discuss the building blocks of \newsreclib{} and provide an overview of the readily available models. For a detailed documentation on the usage of the library, we refer to its project page. 

\section{NewsRecLib -- the Library}

Figure \ref{fig:framework} depicts the structure of \newsreclib{}, comprising different functional modules: 
from data modules for downloading and processing datasets to recommendation modules for training and evaluating a particular NNR. The overall pipeline of an experiment is built automatically from the high-level experimental flow provided by the user in the form of a single Hydra configuration file.

\subsection{Modularization and Extensiblity}

\newsreclib{} is highly modularized: it decouples core components to the largest extent possible. This allows for combinations of different news encoders (e.g., over different input features -- text, aspects, entities) with different user modeling techniques, click fusion strategies, and training objectives. \newsreclib{} is easily extensible with new features: the user only needs to write a new sub-component class (e.g., category encoder), or, in the case of new datasets or recommenders, to define a new PyTorch Lightning data module or (model) module, respectively.

\begin{figure*}[t]
  \centering
  \includegraphics[width=0.75\textwidth]{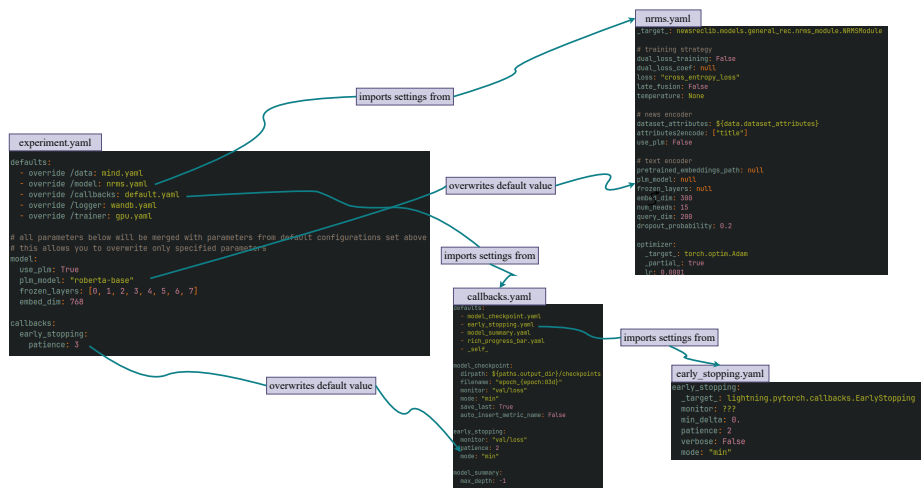}
  \caption{A minimal configuration example for training an NRMS \cite{wu2019nrms} model. All settings defined in the main and the imported configuration files are merged and persisted into a single configuration object.}
  \label{fig:config_example}
\end{figure*}

Concretely, we decouple the essential building blocks of a NNR, namely the \textit{news encoder} (NE), the \textit{user encoder} (UE), and the \textit{click predictor}. NE is further decomposed into a configurable set of feature encoders (i.e., components that embed different aspects of the news, e.g., title, topical category or named entities).
Different model components can be interchanged with corresponding sub-modules of other recommenders, ensuring freedom in choosing each building block of a model independently of the other components (i.e., by mixing the NE of ``NNR 1'' with the UE of ``NNR 2''), in contrast to practices in existing NNR libraries, in which sub-components are tied to concrete NNR architectures that introduced them. Because of this, \newsreclib{} allows for clear-cut and comprehensive analyses of impact of NNR components on their overall performance.\footnote{E.g., we leveraged an earlier version of \newsreclib{} to analyze the impact of click behavior fusion strategies and training objectives on NNRs' performance \cite{iana2023simplifying}.}
\newsreclib{} currently implements feature encoders used in pre-implemented models (see Appendix \S\ref{sec:supported_models}); users can, however, easily incorporate new ones (e.g., an image encoder) by extending the respective class.

\subsection{Configurability and Reproducibility}

Reproducibility strongly relies on the transparency of each step and component in the pipeline, as well as the availability of metadata regarding the factors that influence the model (e.g., hyperparameter values, training objective) and the environment in which it is trained and evaluated (e.g., library versions). 
Because of this, \newsreclib{} leverages the Hydra\footnote{\url{https://hydra.cc/}} framework \cite{Yadan2019Hydra} to decouple the experiment configuration (i.e., a pipeline of modules) from the concrete implementations (i.e., source code) of the modules. 

Each concrete module setting is specified and retrieved automatically from a dedicated configuration file which can be accessed by all the pipeline components. A variety of callbacks supported by PyTorch Lightning (e.g., model checkpointing, early stopping, debugging) can be defined, and modified via a corresponding configuration. A single configuration file guides each experiment: the default configurations of the used modules and callbacks are hierarchically inherited and can be overridden. 
Experiment configurations can also be overwritten directly from the command line, removing the need to store many similar configuration files: this facilitates fast experimentation and minimizes boilerplate code. Experiments can be executed on CPU, GPU, and in a distributed fashion by specifying the type of accelerator supported in PyTorch Lightning. The integration with extensive logging capabilities (see \S \ref{subsec:profiling}) ensures that any modifications are persistently stored in the experiment directory, together with other log files and model checkpoints.  

Fig. \ref{fig:config_example} shows a minimal configuration example for an experiment that trains an instance of the NRMS \cite{wu2019nrms} model. The main configuration file \texttt{experiment.yaml} guides the pipeline. It inherits the data and model-specific configurations from \texttt{mind.yaml} and \texttt{nrms.yaml}, which specify the default configurations of the data module and NNR model, respectively. \texttt{experiment.yaml} further uses the default configurations for the WandB logger, the trainer, and various callbacks. The example also illustrates the interplay between modularization and configurability: we replace the original NE of the NRMS model with a pretrained language model (in this case \texttt{roberta-base}).

\subsection{Performance Evaluation and Profiling}
\label{subsec:profiling}

With Hydra's pluggable architecture as its backbone, every part of the recommendation pipeline is transparent to the user. \newsreclib{} records comprehensive information during training, including number of trainable model parameters and total model size, runtimes, training and validation losses. Moreover, it stores important metadata regarding hyperparameter settings, operating system, PyTorch version, environment details, and dependencies between libraries. Any profiler supported by PyTorch can be incorporated by a simple modification of the corresponding configuration file.

\newsreclib{} supports widely used loggers like WandB\footnote{\url{https://wandb.ai/site}} \cite{wandb} and Tensorboard\footnote{\url{https://www.tensorflow.org/tensorboard}} \cite{abadi2016tensorflow}.
Moreover, users can export evaluation metrics for further analysis. Appendix \ref{sec:logging} shows an example of the logging output. 
We rely on TorchMetrics\footnote{\url{https://torchmetrics.readthedocs.io/en/stable/}} \cite{torchmetrics2022} for model evaluation. Users can track numerous metrics ranging from accuracy-based to beyond-accuracy (e.g., diversity) performance. New metrics can be easily added to the pipeline, either by defining the necessary callbacks in the case of metrics already available in TorchMetrics, or by implementing a custom metric as a subclass of the base \texttt{Metric} class in TorchMetrics.  

\subsection{Hyperparameter Optimization}
\begin{figure}[t]
  \centering
  \includegraphics[width=0.8\columnwidth]{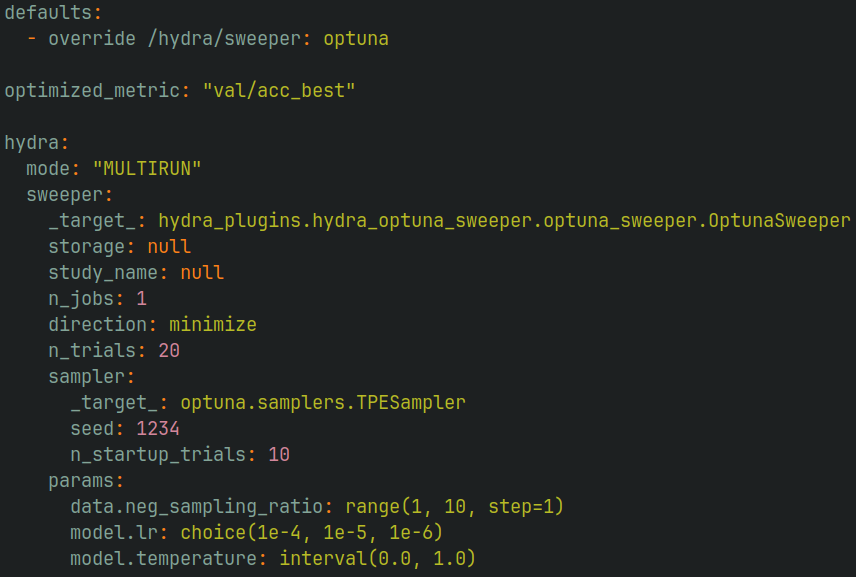}
   \caption{Example of a hyperparameter optimization process. The configuration first runs 10 trials of a search using Bayesian optimization. The hyperparameter search space is defined by indicating the interval, range or choice of values for each desired parameter.}
  \label{fig:hpo_example}
\end{figure}

NNR performance heavily depends on model hyperparameters, making hyperparameter optimization a crucial ingredient in the empirical evaluations of NNRs. \newsreclib{} supports hyperparameter tuning using the Optuna framework \cite{optuna_2019}, which offers a wide range of samplers, such as random search, grid search, and Bayesian optimization \cite{bergstra2011algorithms,ozaki2020multiobjective}.\footnote{\url{https://optuna.readthedocs.io/en/stable/index.html}} 
In conjunction with the modularity of \newsreclib{}, this allows nearly every component of a news recommender to be treated as a hyperparameter, so that users can optimize the choice of encoders or scoring functions. 
Figure \ref{fig:hpo_example} shows a basic multi-objective hyperparameter search 
over the number of negative samples, the model's learning rate, and temperature for the supervised contrastive loss.

\subsection{Available Modules}
\newsreclib{} currently encompasses two popular benchmark datasets, 13 news recommendation models, and various evaluation metrics.

\rparagraph{Datasets.}
We provide out-of-the-box utilities for two prominent monolingual news recommendation benchmarks: MIND \cite{wu2020mind} (with English news) and Adressa \cite{gulla2017adressa} (with Norwegian news). For both datasets, \newsreclib{} supports automatic downloading (when available)\footnote{Note that for the Adressa dataset, only a limited version of the dataset is available for download. For the full version containing additional features, users should contact the authors, as detailed in \url{https://reclab.idi.ntnu.no/dataset/}},
data parsing, and pre-processing functionalities to create a unified PyTorch Lightning datamodule. For both datasets, we include their small and large versions, MINDsmall and MINDlarge, and Adressa-1 week and 10 weeks, respectively. 

Since \citet{wu2020mind} do not publicly release test labels for MIND, we use the provided validation portion for testing, and split the respective training set into temporally disjoint training and validation portions. We follow established practices on splitting the Adressa dataset \cite{hu2020graph,xu2023group} into train, validation, and test sets. In contrast to MIND, which consists of impression log (lists of clicked and non-clicked news by the user), the Adressa dataset contains only positive samples \cite{gulla2017adressa}. Following \citet{yi2021efficient}, we build impressions by randomly sampling 20 news as negatives for each clicked article.

We additionally automatically annotate datasets with sentiment labels obtained by VADER \cite{hutto2014vader}, a monolingual (English) rule-based algorithm (only for MIND), and a multilingual sentiment classification model of \citet{barbieri2021xlm}, fine-tuned from XLM-RoBERTa Base \cite{conneau2020unsupervised}. 

\rparagraph{Recommendation Models.}
\newsreclib{} provides implementations for 10 general-purpose NRRs
and 3 fairness-aware recommenders.
To support analysis of model components, for the models that did not use PLMs in their NEs (but rather contextualized embeddings with convolutional or attention layers), we implement an additional variant with a PLM-based NE (as proposed in \citet{wu2021empowering}). 
Furthermore, models can be trained either with \textit{early fusion}, i.e., learning a parameterized user encoder to aggregate embeddings of news or the simpler \textit{late fusion} strategy proposed in \citet{iana2023simplifying}, which replaces explicit user encoders with parameter-efficient dot products between candidate and clicked news embeddings.
Appendix \ref{sec:supported_models} details all available configurations for each recommendation model.

\rparagraph{Training Objectives.}
Most NNR models are trained with point-wise classification objectives \cite{wang2018dkn,wu2019naml,wu2019nrms} with negative sampling \cite{wu2019npa,wu2021rethinking}. In \citet{iana2023simplifying}, we have shown that contrastive learning constitutes a viable alternative. At the same time, combining point-wise classification with contrastive objectives has been successfully employed in related tasks \cite{gunel2020supervised}. We thus implement three training objectives: cross-entropy loss, supervised contrastive loss \cite{khosla2020supervised}, and a dual objective that is a weighted average between the two.

\rparagraph{Evaluation Metrics.}
\newsreclib{} integrates standard accuracy-based metrics, such as AUC, MRR, and nDCG$@k$. Additionally, we implement aspect-based diversity and aspect-based personalization defined in \citet{iana2023train}. The availability of these beyond-accuracy metrics enables multi-faceted evaluation of NNRs. 

\section{Comparison to Related Frameworks}

In the past decade, numerous frameworks for the development and comprehensive evaluation of recommender systems have been proposed to address the problem of reproducibility in the field \cite{gantner2011mymedialite,ekstrand2011rethinking,ekstrand2020lenskit,guo2015librec,kula2017spotlight,da2018case,salah2020cornac,hug2020surprise,sun2020we,anelli2021elliot}. News recommendation poses different challenges for practitioners in comparison to recommendation in domains such as movies, music, or e-commerce \cite{raza2022news,iana2022survey}. However, few of the existing and widely used libraries offer support for news recommenders, and especially for the modern neural news recommendation models.

Microsoft Recommenders \cite{graham2019microsoft,argyriou2020microsoft} and RecBole \cite{zhao2021recbole,zhao2022recbole} provide implementation for five and three NNRs, respectively, as well as utilities for the MIND dataset. Nonetheless, other datasets, more recent approaches, and in particular fairness-aware models and beyond-accuracy metrics 
are not supported. StreamingRec \cite{jugovac2018streamingrec} is a framework for evaluating streaming-based news recommenders, covering a wide range of algorithms, from trivial baselines (e.g., recently published, most popular) 
or item-to-item based collaborative filtering or session-based nearest neighbor techniques, to association rule methods and content-based approaches. However, it does not support any of the recent neural models. In these libraries, the sub-modules of a specific recommender are not decoupled from the overall model, which impedes experimentation with and analysis of different model components and training strategies. 

In contrast to these frameworks, \newsreclib{} focuses solely on the state-of-the-art neural news recommendation models, providing utilities for the most used benchmark datasets, architectures, training techniques, and evaluation metrics tailored to news recommendation. 
\newsreclib{} unifies disparate implementations of recent neural news recommenders in a single open-source library that is built on top of mature frameworks for deep learning (PyTorch Lightning), evaluation (TorchMetrics), and configuration (Hydra). 
\section{Experiments}

\begin{table*}
    \centering

    \resizebox{\textwidth}{!}{%
        \begin{tabular}{cl|cccc:cc|cccc:cc}
        \hline
        
        & \multicolumn{1}{c}{}
        & \multicolumn{6}{c}{\textbf{MIND}} 
        & \multicolumn{6}{c}{\textbf{Adressa}} \\  
        \cmidrule(lr){3-10} \cmidrule(lr){9-14}
    
        & Model
        & AUC 
        &  \cellcolor{Gray} MRR
        & nDCG@5
        &  \cellcolor{Gray} nDCG@10
        & D\textsubscript{ctg}@10  
        &  \cellcolor{Gray} D\textsubscript{snt}@10 
        
        & AUC 
        &  \cellcolor{Gray} MRR
        & nDCG@5
        &  \cellcolor{Gray} nDCG@10
        & D\textsubscript{ctg}@10  
        &  \cellcolor{Gray} D\textsubscript{snt}@10 
        \\
        \hline

        & DKN
        & 50.0$\pm$0.0 
        &  \cellcolor{Gray} 26.3$\pm$0.4 
        & 24.6$\pm$0.5 
        &  \cellcolor{Gray} 31.5$\pm$0.3 
        & 50.4$\pm$1.0 
        &  \cellcolor{Gray} 66.0$\pm$0.6 

        & -- 
        &  \cellcolor{Gray} -- 
        & -- 
        &  \cellcolor{Gray} -- 
        & -- 
        & \cellcolor{Gray}  -- 
        \\ 

        & NPA
        & 55.1$\pm$0.6 
        &  \cellcolor{Gray} 28.5$\pm$1.1 
        & 26.4$\pm$1.1 
        &  \cellcolor{Gray} 32.9$\pm$1.0 
        & 51.8$\pm$0.2 
        &  \cellcolor{Gray} 67.5$\pm$0.7 

        & 53.3$\pm$3.5 
        &  \cellcolor{Gray} 31.8$\pm$1.9 
        & 30.4$\pm$0.3 
        &  \cellcolor{Gray} 38.2$\pm$2.8 
        & 31.6$\pm$0.3 
        &  \cellcolor{Gray} 60.7$\pm$0.4 
        \\ 

        & NRMS
        & 54.1$\pm$0.8 
        &  \cellcolor{Gray} 27.2$\pm$0.6 
        & 25.3$\pm$0.5 
        &  \cellcolor{Gray} 31.9$\pm$0.4 
        & 52.1$\pm$0.6 
        &  \cellcolor{Gray} 65.9$\pm$1.7 

        & 63.8$\pm$4.7 
        &  \cellcolor{Gray} 30.5$\pm$3.6 
        & 28.6$\pm$5.5 
        &  \cellcolor{Gray} 37.1$\pm$4.8 
        & 31.7$\pm$0.3 
        &  \cellcolor{Gray} 60.7$\pm$0.7 
        \\ 

        & NAML
        & 50.2$\pm$0.0 
        &  \cellcolor{Gray} \underline{33.4$\pm$0.6} 
        & \underline{31.8$\pm$0.7} 
        &  \cellcolor{Gray} \underline{38.1$\pm$0.5} 
        & 47.0$\pm$1.0 
        &  \cellcolor{Gray} 66.9$\pm$0.3 

        & 50.0$\pm$0.0 
        &  \cellcolor{Gray} 37.8$\pm$3.5 
        & 38.2$\pm$4.1 
        &  \cellcolor{Gray} 45.1$\pm$3.6 
        & 31.5$\pm$4.6 
        &  \cellcolor{Gray} 60.7$\pm$0.4 
        \\ 
    
        & LSTUR\footnotemark
        & 58.8$\pm$2.1 
        &  \cellcolor{Gray} 32.2$\pm$0.9 
        & 30.4$\pm$0.9 
        &  \cellcolor{Gray} 36.8$\pm$0.9 
        & 43.1$\pm$1.2 
        &  \cellcolor{Gray} 65.6$\pm$0.6 

        & 68.1$\pm$2.4 
        &  \cellcolor{Gray} \textbf{38.0$\pm$1.7} 
        & \textbf{39.1$\pm$2.3} 
        &  \cellcolor{Gray} \textbf{45.9$\pm$2.4} 
        & 27.7$\pm$2.4 
        &  \cellcolor{Gray} 60.1$\pm$0.3 
        \\ 
    
        & TANR
        & 53.0$\pm$4.1 
        &  \cellcolor{Gray} 30.7$\pm$0.6 
        & 29.0$\pm$0.5 
        &  \cellcolor{Gray} 35.3$\pm$0.4 
        & 50.5$\pm$0.4 
        &  \cellcolor{Gray} 66.7$\pm$0.8 

        & 50.3$\pm$0.5 
        &  \cellcolor{Gray} 33.4$\pm$3.4 
        & 32.9$\pm$4.7 
        &  \cellcolor{Gray} 40.0$\pm$4.1 
        & 29.8$\pm$0.9 
        &  \cellcolor{Gray} 60.1$\pm$0.3 
        \\ 

        & CAUM
        & 59.5$\pm$0.6 
        &  \cellcolor{Gray}  \cellcolor{Gray} 33.1$\pm$0.4 
        & 31.2$\pm$0.5 
        &  \cellcolor{Gray} 37.7$\pm$0.5 
        & 47.4$\pm$0.5 
        &  \cellcolor{Gray} 66.7$\pm$0.8 

        & \underline{72.5$\pm$2.3} 
        &  \cellcolor{Gray} 36.0$\pm$3.1 
        & 37.7$\pm$4.4 
        &  \cellcolor{Gray} 44.9$\pm$3.1 
        & 29.3$\pm$3.3 
        &  \cellcolor{Gray} 60.5$\pm$0.3 
        \\ 
    
        & MINS
        & 56.1$\pm$1.5 
        &  \cellcolor{Gray} 31.0$\pm$1.5 
        & 29.4$\pm$1.5 
        &  \cellcolor{Gray} 35.7$\pm$1.5 
        & 47.0$\pm$1.7 
        &  \cellcolor{Gray} 67.6$\pm$1.1 

        & \textbf{73.8$\pm$3.2} 
        &  \cellcolor{Gray} \underline{37.4$\pm$2.5} 
        & \underline{38.8$\pm$4.1} 
        &  \cellcolor{Gray} \underline{45.8$\pm$3.2} 
        & 32.4$\pm$0.8 
        &  \cellcolor{Gray} 60.6$\pm$0.3 
        \\ 

        \multirow{-9}{*}{\rotatebox[origin=c]{0}{\textbf{GeneralRec}}}
        & CenNewsRec
        & 54.7$\pm$1.3 
        &  \cellcolor{Gray} 26.9$\pm$0.8 
        & 25.4$\pm$0.8 
        &  \cellcolor{Gray} 32.0$\pm$0.7 
        & 50.9$\pm$0.7 
        & \cellcolor{Gray}  68.1$\pm$0.7 

        & 62.3$\pm$2.1 
        &  \cellcolor{Gray} 29.3$\pm$2.6 
        & 26.9$\pm$3.9 
        &  \cellcolor{Gray} 35.1$\pm$3.2 
        & 31.7$\pm$0.5 
        &  \cellcolor{Gray} 60.7$\pm$0.3 
        \\ 

        \hdashline
        
        & SentiRec
        & 52.0$\pm$0.5 
        &  \cellcolor{Gray} 27.2$\pm$0.9 
        & 25.2$\pm$1.0 
        &  \cellcolor{Gray} 31.8$\pm$0.8 
        & 52.5$\pm$1.2 
        &  \cellcolor{Gray} 67.7$\pm$1.1 

        & 55.0$\pm$0.7 
        &  \cellcolor{Gray} 26.9$\pm$0.4 
        & 24.3$\pm$0.7 
        &  \cellcolor{Gray} 30.1$\pm$0.7 
        & \underline{35.2$\pm$0.1} 
        &  \cellcolor{Gray} \textbf{66.1$\pm$0.7} 
        \\ 

        \multirow{-2}{*}{\rotatebox[origin=c]{0}{\textbf{FairRec}}}
        & SentiDebias
        & 56.6$\pm$1.7 
        &  \cellcolor{Gray} 25.4$\pm$0.7 
        & 23.7$\pm$0.9 
        &  \cellcolor{Gray} 30.3$\pm$0.6 
        & 53.5$\pm$1.3 
        &  \cellcolor{Gray} \underline{68.1$\pm$1.3} 

        & 66.5$\pm$0.9 
        &  \cellcolor{Gray} 29.4$\pm$0.7 
        & 29.2$\pm$1.6 
        &  \cellcolor{Gray} 36.9$\pm$1.2 
        & 31.3$\pm$0.8 
        &  \cellcolor{Gray} 61.1$\pm$0.3 
        \\ 

        \hline

        & NRMS-PLM
        & 50.0$\pm$0.0 
        &  \cellcolor{Gray} 21.9$\pm$2.8 
        & 19.5$\pm$2.9 
        &  \cellcolor{Gray} 26.0$\pm$3.0 
        & 53.2$\pm$1.7 
        &  \cellcolor{Gray} 66.1$\pm$3.4 
       
        & 53.1$\pm$2.7 
        &  \cellcolor{Gray} 34.9$\pm$2.5 
        & 34.7$\pm$3.0 
        &  \cellcolor{Gray} 42.8$\pm$2.8 
        & 32.3$\pm$1.2 
        &  \cellcolor{Gray} 61.6$\pm$0.3 
        \\ 

        & NAML-PLM
        & 52.8$\pm$2.4 
        &  \cellcolor{Gray} 30.0$\pm$1.2 
        & 28.2$\pm$1.3 
        &  \cellcolor{Gray} 34.7$\pm$1.2 
        & 39.3$\pm$2.5 
        &  \cellcolor{Gray} 66.9$\pm$0.6 

        & 50.0$\pm$0.0 
        &  \cellcolor{Gray} 35.3$\pm$2.8 
        & 35.0$\pm$3.8 
        &  \cellcolor{Gray} 41.3$\pm$3.7 
        & 26.7$\pm$6.6 
        &  \cellcolor{Gray} 60.6$\pm$0.5 
        \\ 
    
        & LSTUR-PLM
        & 50.0$\pm$0.0 
        &  \cellcolor{Gray} 30.7$\pm$0.6 
        & 29.0$\pm$0.6 
        &  \cellcolor{Gray} 35.3$\pm$0.6 
        & 36.6$\pm$0.9 
        &  \cellcolor{Gray} 67.0$\pm$0.9 

         & 55.5$\pm$2.3 
        &  \cellcolor{Gray} 30.4$\pm$1.6 
        & 28.8$\pm$2.3 
        &  \cellcolor{Gray} 35.3$\pm$2.2 
        & 22.3$\pm$2.8 
        &  \cellcolor{Gray} 60.9$\pm$0.4 
        \\ 
    
        & TANR-PLM
        & 50.0$\pm$0.7 
        &  \cellcolor{Gray} 25.9$\pm$3.5 
        & 23.3$\pm$3.6 
        &  \cellcolor{Gray} 29.8$\pm$3.4 
        & 47.6$\pm$6.8 
        &  \cellcolor{Gray} 61.4$\pm$3.0 

        & 50.0$\pm$0.0 
        &  \cellcolor{Gray} 35.5$\pm$3.8 
        & 35.1$\pm$5.1 
        &  \cellcolor{Gray} 41.8$\pm$4.7 
        & 24.8$\pm$10.0 
        &  \cellcolor{Gray} 59.9$\pm$1.0 
        \\ 

        & CAUM-PLM
        & \underline{59.7$\pm$2.0} 
        &  \cellcolor{Gray} 32.8$\pm$0.5 
        & 31.0$\pm$0.6 
        &  \cellcolor{Gray} 37.2$\pm$0.5 
        & 44.3$\pm$2.2 
        & \cellcolor{Gray}  67.5$\pm$0.8 

        & 66.1$\pm$2.3 
        &  \cellcolor{Gray} 30.7$\pm$68 
        & 30.6$\pm$8.4 
        &  \cellcolor{Gray} 35.7$\pm$9.9 
        & 22.9$\pm$3.4 
        &  \cellcolor{Gray} 60.4$\pm$0.4 
        \\ 
    
        & MINS-PLM
        & 50.0$\pm$0.7 
        &  \cellcolor{Gray} 22.4$\pm$3.5 
        & 20.2$\pm$3.9 
        &  \cellcolor{Gray} 26.5$\pm$4.0 
        & 50.6$\pm$3.2 
        &  \cellcolor{Gray} 67.3$\pm$1.1 

        & 65.3$\pm$4.4 
        &  \cellcolor{Gray} 33.1$\pm$2.8 
        & 31.5$\pm$4.6 
        &  \cellcolor{Gray} 40.4$\pm$3.9 
        & 26.6$\pm$5.5 
        &  \cellcolor{Gray} 60.5$\pm$0.5 
        \\ 

        & CenNewsRec-PLM
        & 50.0$\pm$0.2 
        &  \cellcolor{Gray} 21.2$\pm$2.8 
        & 18.9$\pm$2.9 
        &  \cellcolor{Gray} 25.4$\pm$2.8 
        & \underline{54.2$\pm$1.3} 
        &  \cellcolor{Gray} 67.0$\pm$1.7 

        & 54.4$\pm$5.3 
        &  \cellcolor{Gray} 35.8$\pm$3.1 
        & 35.9$\pm$3.3 
        &  \cellcolor{Gray} 42.8$\pm$2.1 
        & 31.6$\pm$0.8 
        &  \cellcolor{Gray} 61.0$\pm$0.6 
        \\ 

        \multirow{-8}{*}{\rotatebox[origin=c]{0}{\textbf{GeneralRec}}}
        & MINER\footnotemark{}
        & 51.2$\pm$0.4 
        &  \cellcolor{Gray} 24.2$\pm$0.5 
        & 22.0$\pm$0.6 
        &  \cellcolor{Gray} 28.2$\pm$0.5 
        & \textbf{54.8$\pm$0.3} 
        &  \cellcolor{Gray} 68.8$\pm$0.6 

        & 55.3$\pm$6.9 
        &  \cellcolor{Gray} 33.5$\pm$2.2 
        & 33.1$\pm$3.3 
        &  \cellcolor{Gray} 39.1$\pm$3.3 
        & 32.4$\pm$1.4 
        &  \cellcolor{Gray} 61.2$\pm$1.4
        \\ 

        \hdashline

        & SentiRec-PLM
        & 50.0$\pm$0.6 
        &  \cellcolor{Gray} 24.7$\pm$0.7 
        & 22.6$\pm$0.6 
        &  \cellcolor{Gray} 29.1$\pm$0.6 
        & 52.3$\pm$2.4 
        &  \cellcolor{Gray} 67.2$\pm$2.1 

        & 61.2$\pm$3.0 
        &  \cellcolor{Gray} 31.6$\pm$3.4 
        & 30.4$\pm$4.4 
        &  \cellcolor{Gray} 38.2$\pm$4.4 
        & 32.9$\pm$1.7 
        &  \cellcolor{Gray} 59.9$\pm$2.4 
        \\ 

        & SentiDebias-PLM
        & 51.0$\pm$0.5 
        &  \cellcolor{Gray} 28.7$\pm$0.4 
        & 27.5$\pm$0.4 
        &  \cellcolor{Gray} 34.0$\pm$0.4 
        & 47.7$\pm$2.0 
        &  \cellcolor{Gray} 67.9$\pm$1.7 

        & 67.3$\pm$2.8 
        &  \cellcolor{Gray} 37.1$\pm$3.6 
        & 38.0$\pm$5.1 
        &  \cellcolor{Gray} 45.3$\pm$3.8 
        & 32.6$\pm$1.2 
        &  \cellcolor{Gray} 61.5$\pm$1.0 
        \\

        \multirow{-3}{*}{\rotatebox[origin=c]{0}{\textbf{FairRec}}}
        & MANNeR\footnotemark{}
        & \textbf{66.2$\pm$1.0} 
        &  \cellcolor{Gray} \textbf{36.7$\pm$1.3} 
        & \textbf{35.1$\pm$1.3} 
        &  \cellcolor{Gray} \textbf{41.1$\pm$1.1} 
        & 50.5$\pm$0.3 
        &  \cellcolor{Gray} \textbf{68.2$\pm$0.4} 

         & 67.6$\pm$4.3 
        &  \cellcolor{Gray} 31.9$\pm$2.8 
        & 30.5$\pm$4.1 
        &  \cellcolor{Gray} 38.9$\pm$3.9 
        & \textbf{39.2$\pm$0.4} 
        &  \cellcolor{Gray} \underline{64.9$\pm$0.5} 
        \\
       
        \hline
        \end{tabular}%
    }    
    \caption{Recommendation and aspectual diversity (in terms of topical categories $D_{ctg}$ and sentiments $D_{snt}$) performance of different neural news recommenders. We report averages and standard deviations across five different runs. The best results per column are highlighted in bold, the second best are underlined. The dashed line separates the general (\texttt{GeneralRec}) from the fairness-aware  (\texttt{FairRec}) recommendation models.}
    \label{tab:benchmark_results}
\end{table*}

\addtocounter{footnote}{-3}
\stepcounter{footnote}\footnotetext{We use the LSTUR\textsubscript{ini} version of the model. For details, refer to \citet{an2019neural}.}
\stepcounter{footnote}\footnotetext{We use the MINER \textit{weighted} version of the model. For details, refer to \citet{li2022miner}.}
\stepcounter{footnote}\footnotetext{We use the MANNeR version which performs multi-aspect diversification with $\lambda_{ctg}=-0.15$ and $\lambda_{snt}=-0.25$ for MIND, and $\lambda_{ctg}=-0.35$ and $\lambda_{snt}=-0.25$ for Adressa, respectively. For details, refer to \citet{iana2023train}.}

We conduct experiments with the pre-implemented recommendation models from \newsreclib{} to investigate their performance when (1) trained with the original architecture (e.g., NE based on word embeddings and contextualization layer) and (2) trained with a PLM-based NE \cite{wu2021empowering}. 

\subsection{Datasets and Experimental Setup}

We carry out the evaluation on the MINDsmall \cite{wu2020mind} (denoted MIND) and Adressa-1 week (denoted Adressa) \cite{gulla2017adressa} benchmark datasets. We evaluate two versions of the models, namely (1) with the original NE and (2) the NE modified to use a PLM \cite{wu2021empowering} (if not used in the original NE). We use RoBERTa Base \cite{liu2019roberta} and NB-BERT Base \cite{kummervold2021operationalizing,nielsen-2023-scandeval} for experiments on MIND and Adressa, respectively. In both cases, we fine-tune only the last four layers of the PLM in the interest of computational efficiency. We use 100-dimensional TransE embeddings \cite{bordes2013translating} pretrained on Wikidata as input to the entity encoder for models using named entities as input features to their NEs, a maximum history length of 50, and set all other model-specific hyperparameters to optimal values reported in the respective papers. We train all models with mixed precision, and optimize with the Adam algorithm \cite{kingma2014adam}, with the learning rate of 1e-4. We train models with a PLM-empowered NE for 10 epochs, and the model variant without PLMs for 20 epochs. Since Adressa contains no abstract or disambiguated named entities, we use only the title for the models benchmarked on that dataset.

\subsection{Results}
Table \ref{tab:benchmark_results} summarizes the results on content-based recommendation performance (w.r.t. AUC, MRR, nDCG@5, nDCG@10) and aspect diversification for topical categories ($D_{ctg}$) and sentiment ($D_{snt}$), as per \citet{iana2023train}. 
We find that PLM-based NEs do not necessarily lead to performance improvements. We hypothesize that this is due to the dataset size: a PLM-based NE requires training a larger number of parameters than one which contextualizes pretrained word embeddings with a CNN or attention network. Note that rather small improvements of PLM-empowered NEs over original NEs have been shown only for larger-scale datasets \cite{wu2021empowering}. These findings indicate that more research is needed to understand under which settings older NEs can still benefit NNRs. 
%
MANNeR, with its late click behavior fusion approach, outperforms all other models on MIND, but it underperforms on Adressa. Note that the contrastive learning training approach adopted by MANNeR \cite{iana2023train} benefits from larger training datasets, and MINDsmall has roughly five times as many news as Adressa 1-week. Expectedly, w.r.t. aspect-based diversity, NNRs with diversification objectives (e.g., for sentiment) outperform models trained only to maximize content-based accuracy. 
\section{Conclusion}

In this work, we introduced \newsreclib{}, a highly configurable, modular and easily extensible framework for neural news recommendation. Our library is specifically designed to foster reproducible research in recommender systems and rigorous evaluation of models -- users only need to create a single configuration file for an experiment. We briefly described the underlying principles of \newsreclib{} and the structure of its building blocks. The framework currently provides two standard benchmark datasets, loading and pre-processing functions, 13 neural recommendation models, different training objectives and hyperparameters optimization strategies, numerous evaluation metrics, extensive logging capabilities, and GPU support. 
We believe that \newsreclib{} is a useful tool for the community that will (i) catalyze reproducible NNR research, (ii) foster fairer comparisons between the models, and (iii) facilitate identification of NNR components that drive their performance.

\section*{Limitations}

While we have striven to build a comprehensive library for the design and fair evaluation of neural news recommendation models, several additional factors must be taken into account. Firstly, even though we aim to replicate the original implementations of the models to the highest degree possible, discrepancies in our code and results can arise from the usage of different frameworks, as well as scarce availability of implementation details in the source code or publications of some of the recommenders. Secondly, our library is heavily dependent on the changes and maintenance of the frameworks on which it is built, namely PyTorch Lightning (and by extension, PyTorch), Hydra, TorchMetrics, Optuna. As such, new plugins for logging (e.g., Neptune \cite{neptune_2019}, Comet \cite{rei2020comet}, MLFlow \cite{zaharia2018accelerating}) or hyperparamter optimization (e.g., Ax\footnote{\url{https://ax.dev/}}) need to be integrated with PyTorch Lightning and Hydra. 

Moreover, we rely on open benchmark news datasets for training and evaluating the recommenders. Consequently, any biases that might be contained in the news and user data could be propagated through the recommendation pipeline. Additionally, the usage of these datasets is intertwined with their public availability. Any changes to the datasets or access restrictions are likely to impact the way pre-implemented models in \newsreclib{} can be trained and benchmarked.

Lastly, neural news recommendation is a computationally expensive endeavor which requires availability of large compute resources. Although \newsreclib{} technically supports execution of experiments on CPU, this would be not only highly inefficient and time-consuming, but also infeasible for large-scale datasets with hundreds of thousands of users and news. Consequently, users should ideally have access to GPUs to efficiently use our library. 

\section*{Ethics Statement}
Users of our library should differentiate the recommendation models available in \newsreclib{} from the originals. Consequently, they should explicitly credit and cite both \newsreclib{}, as well as the original implementations, as specified on our GitHub page.  

\section*{Acknowledgements}
The authors acknowledge support by the state of Baden-Württemberg through bwHPC
and the German Research Foundation (DFG) through grant INST 35/1597-1 FUGG.

\bibliography{anthology,custom}
\bibliographystyle{acl_natbib}

\appendix
\section{Logging}
\label{sec:logging}

\subsection{Configuration Logging}
\label{subsec:config_logging}

\begin{figure*}[t]
     \centering
     \begin{subfigure}[b]{\textwidth}
          \centering
          \includegraphics[width=\textwidth]{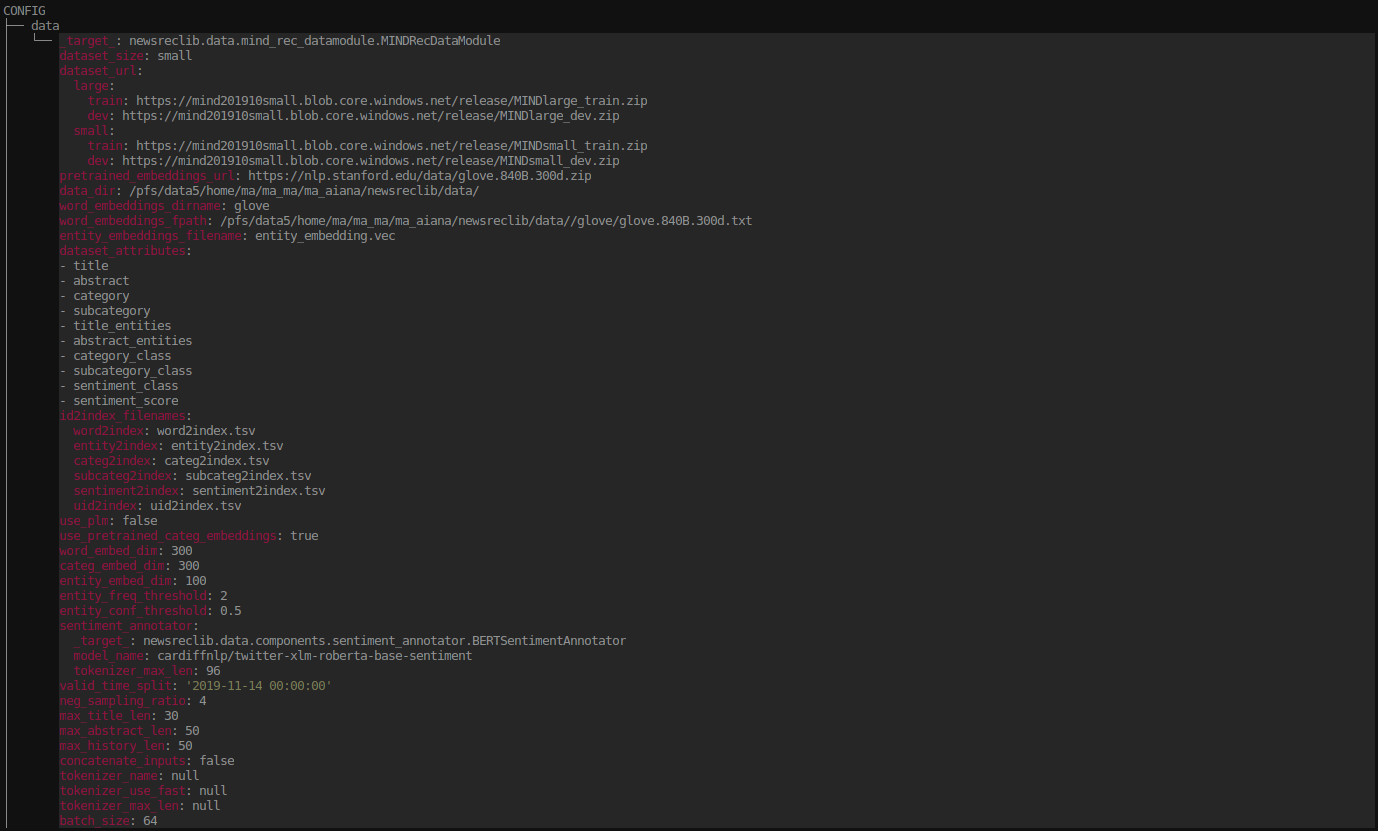}
          \caption{Data module configuration.}
          \label{fig:config_data}
     \end{subfigure}
     \vfill
     \begin{subfigure}[b]{\textwidth}
          \centering
          \includegraphics[width=\textwidth]{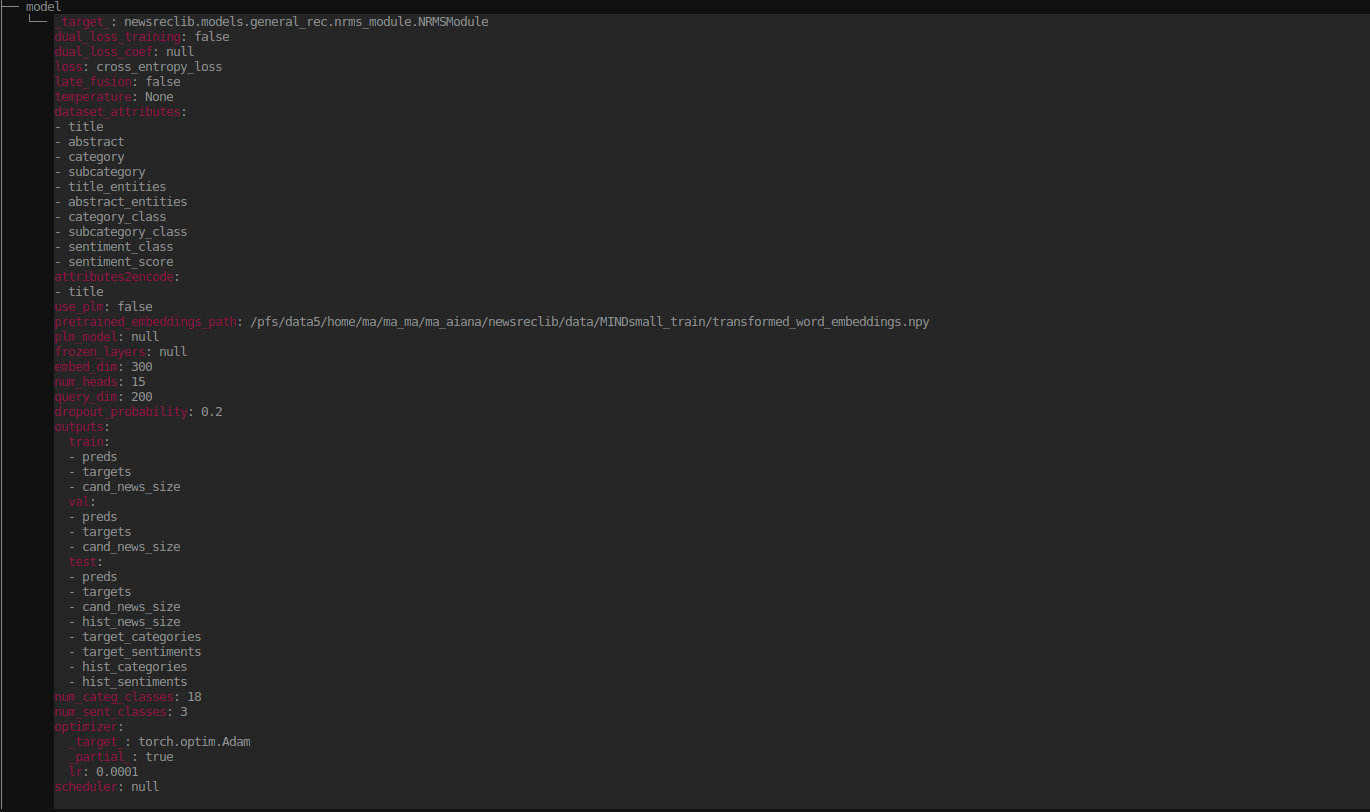}
          \caption{Recommendation module configuration.}
          \label{fig:config_model}
     \end{subfigure}     
     \caption{Example for logging the configurations of the data and the recommendation modules.}
     \label{fig:logging_part1}
\end{figure*}

\begin{figure*}[t]
     \centering
     \begin{subfigure}[b]{\textwidth}
          \centering
          \includegraphics[width=\textwidth]{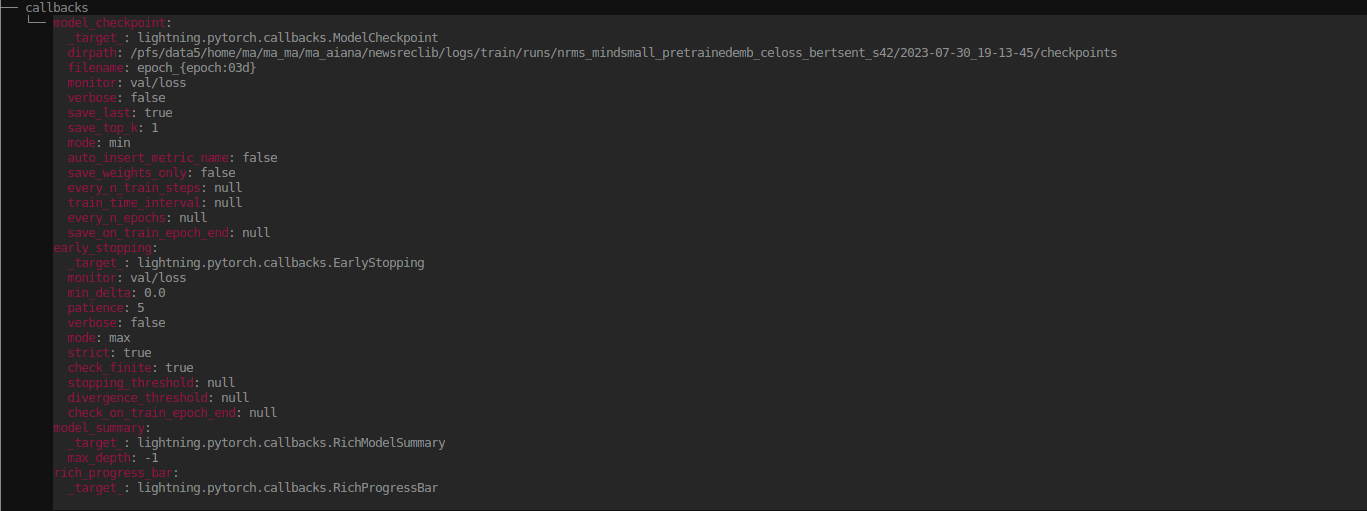}
          \caption{Callbacks configuration.}
          \label{fig:config_callbacks}
     \end{subfigure}
     \vfill
     \begin{subfigure}[b]{\textwidth}
          \centering
          \includegraphics[width=\textwidth]{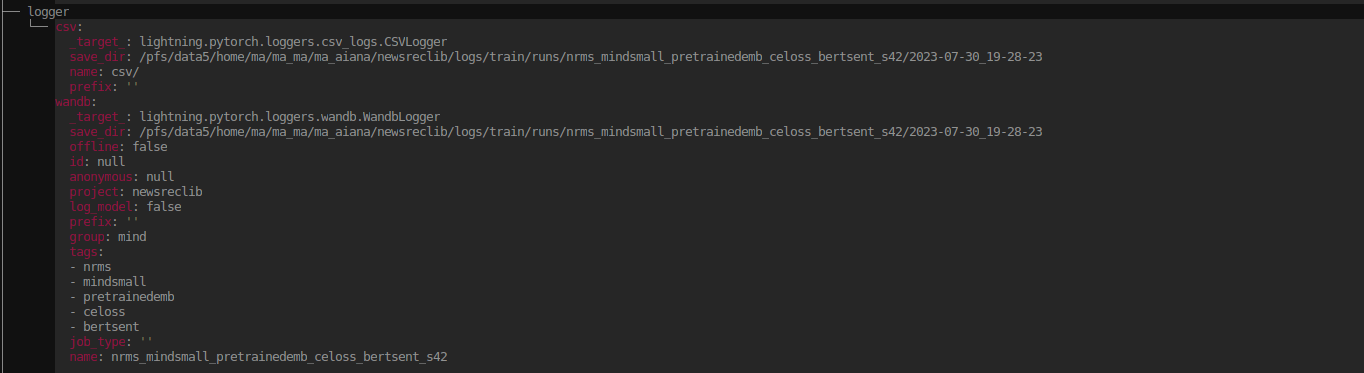}
          \caption{Logger configuration.}
          \label{fig:config_logger}
     \end{subfigure}
     \vfill
     \begin{subfigure}[b]{\textwidth}
          \centering
          \includegraphics[width=\textwidth]{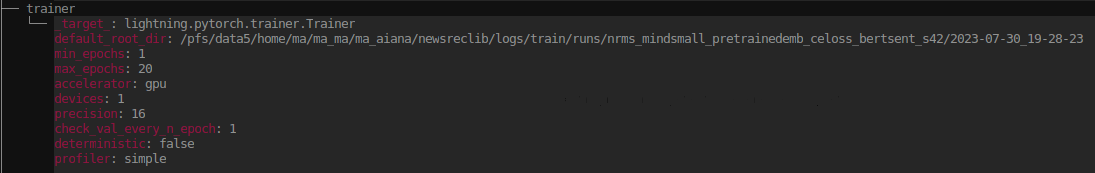}
          \caption{Trainer configuration.}
          \label{fig:config_trainer}
     \end{subfigure}
     \vfill
     \begin{subfigure}[b]{\textwidth}
          \centering
          \includegraphics[width=\textwidth]{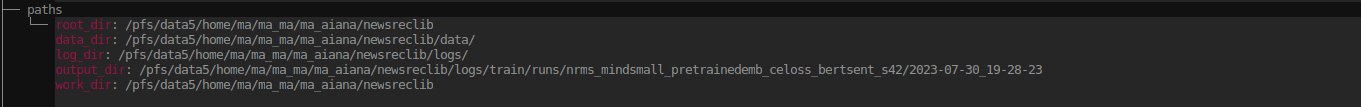}
          \caption{Paths configuration.}
          \label{fig:config_paths}
     \end{subfigure}
        \caption{Example for logging the configurations of the callbacks, loggers, and trainer.}
        \label{fig:logging_part2}
\end{figure*}

Figs. \ref{fig:logging_part1} and \ref{fig:logging_part2} illustrate an example of how the configuration of each of the pipeline's components is logged when the training process is initiated.

\subsection{Model Metadata Logging}
\label{subsec:model_logging}

\begin{figure*}[t]
  \centering
  \includegraphics[width=\textwidth]{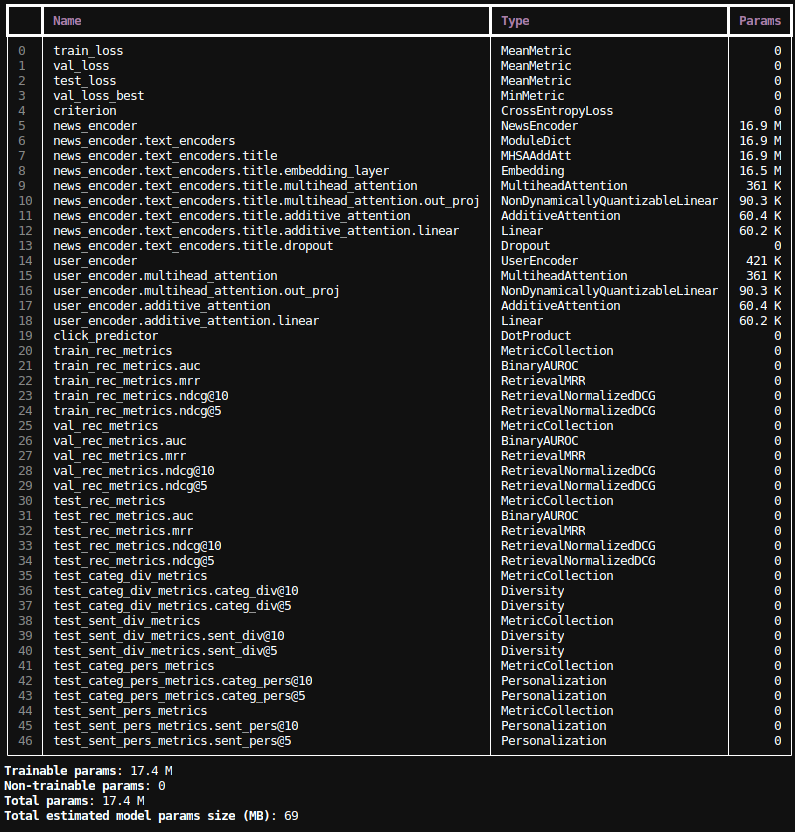}
  \caption{Example for logging the model size, number of trainable and non-trainable model parameters.}
  \label{fig:logging_model_size}
\end{figure*}

Fig. \ref{fig:logging_model_size} shows an example of logging relevant metadata information regarding a model's size and number of parameters.

\section{Supported Recommendation Models and Configurations}
\label{sec:supported_models}

\newsreclib{} provides, to date, implementations of 10 general NNRs:
\begin{itemize}
    \item \textit{DKN} \cite{wang2018dkn} uses a word-entity aligned knowledge-aware convolutional neural network (CNN) \cite{kim2014convolutional} to produce news embeddings. It learns candidate-aware representations of users as the weighted sum of their clicked news embeddings, where the weights are computed by an attention network that takes as input the embeddings of the candidate and of the clicked news. 
    
    \item \textit{NPA} \cite{wu2019npa} contextualizes pretrained word embeddings with a CNN, followed by a personalized attention module. Its UE consists of a similar personalized attention module which aggregates the representations of the users’ clicked news, with projected embeddings of the users IDs as attention queries.
    
    \item \textit{NAML} \cite{wu2019naml} uses a sequence of CNN and additive attention \cite{bahdanau2015neural} to contextualize pretrained word embeddings in its NE. Additionally, it leverages category information, with categories embedded through a linear layer. User representations are learned from the embeddings of users' clicked news with another additive attention layer.
    
    \item \textit{NRMS} \cite{wu2019nrms} learns news representations from pretrained word embeddings and a combination of multi-head self-attention \cite{vaswani2017attention} and additive attention; it embeds users with a two-layer encoder consisting also of multi-head self-attention and additive attention.
    
    \item \textit{LSTUR} \cite{an2019neural} embeds news similarly to NAML \cite{wu2019naml}. However, it learns user representations via recurrent neural networks: it produces short-term user embeddings from the clicked news with a GRU \cite{cho2014learning}, which it combines with a long-term embedding, consisting of a randomly initialized and fine-tuned part.
    
    \item \textit{TANR} \cite{wu2019tanr} injects information on topical categories, by jointly optimizing the NNR for content personalization and topic classification. It uses the same UE and NE architecture as NAML \cite{wu2019naml}, but does not embed categories.
    
    \item \textit{CAUM} \cite{qi2022news} uses a NRMS-based NE, and additionally encodes title entities with attention layers. Moreover, its candidate-aware UE combines a candidate-aware self-attention network which models long-range dependencies between clicked news, conditioned on the candidate, with a candidate-aware CNN that captures short-term user interests from adjacent clicks, again conditioned on the candidate’s content.
    
    \item \textit{MINS} \cite{wang2022news} embeds textual features of news (i.e., title, abstract) in the same manner as NRMS \cite{wu2019nrms}, and categories through a linear embedding layer. Moreover, it uses a combination of multi-head self-attention, multi-channel GRU-based recurrent network, and additive attention to encode users.
    
    \item \textit{CenNewsRec} \cite{qi2020privacy} combines a CNN network with multi-head self-attention and additive attention modules to produce contextualized representations of news. Its UE resembles that of LSTUR \cite{an2019neural}, but it learns long-term user vectors from clicked news using a sequence of multi-head self-attention and attentive pooling networks, as opposed to storing an explicit embedding per user.
    
    \item \textit{MINER} \cite{li2022miner} uses a pretrained BERT \cite{devlin2019bert} model as NE. Its UE learns multiple user representation vectors using a poly attention scheme that extracts interests vectors through additive attention layers.
\end{itemize}

\begin{table*}[t]
    \centering

    \resizebox{\textwidth}{!}{%
        \begin{tabular}{cl|cc|cc|ccc}
        \hline
        & &
        \multicolumn{2}{c}{\textbf{News Encoder}} & 
        \multicolumn{2}{c}{\textbf{Click Behavior Fusion}} & 
        \multicolumn{3}{c}{\textbf{Training Objective}} \\  
        \cmidrule(lr){3-4} \cmidrule(lr){5-6} \cmidrule(lr){7-9}

        & Model
        & Word emb. + contextualization
        & PLM
        & EF
        & LF  
        & CE
        & SCL
        & Dual
        \\
        \hline

        & DKN \cite{wang2018dkn}
        & \Checkmark 
        & \XSolidBrush 
        & \Checkmark 
        & \Checkmark 
        & \Checkmark 
        & \Checkmark 
        & \Checkmark 
        \\

        & \cellcolor{Gray} NPA \cite{wu2019npa}
        & \cellcolor{Gray} \Checkmark 
        & \cellcolor{Gray} \XSolidBrush 
        & \cellcolor{Gray} \Checkmark 
        & \cellcolor{Gray} \Checkmark 
        & \cellcolor{Gray} \Checkmark 
        & \cellcolor{Gray} \Checkmark 
        & \cellcolor{Gray} \Checkmark 
        \\

        & NRMS \cite{wu2019nrms}
        & \Checkmark 
        & \Checkmark 
        & \Checkmark 
        & \Checkmark 
        & \Checkmark 
        & \Checkmark 
        & \Checkmark 
        \\     

        & \cellcolor{Gray} NAML \cite{wu2019naml}
        & \cellcolor{Gray} \Checkmark 
        & \cellcolor{Gray} \Checkmark 
        & \cellcolor{Gray} \Checkmark 
        & \cellcolor{Gray} \Checkmark 
        & \cellcolor{Gray} \Checkmark 
        & \cellcolor{Gray} \Checkmark 
        & \cellcolor{Gray} \Checkmark 
        \\
       
        & LSTUR \cite{an2019neural}
        & \Checkmark 
        & \Checkmark 
        & \Checkmark 
        & \Checkmark 
        & \Checkmark 
        & \Checkmark 
        & \Checkmark 
        \\

        & \cellcolor{Gray} TANR \cite{wu2019tanr}
        & \cellcolor{Gray} \Checkmark 
        & \cellcolor{Gray} \Checkmark 
        & \cellcolor{Gray} \Checkmark 
        & \cellcolor{Gray} \Checkmark 
        & \cellcolor{Gray} \Checkmark 
        & \cellcolor{Gray} \Checkmark 
        & \cellcolor{Gray} \Checkmark 
        \\
       
        & CAUM \cite{qi2022news}
        & \Checkmark 
        & \Checkmark 
        & \Checkmark 
        & \Checkmark 
        & \Checkmark 
        & \Checkmark 
        & \Checkmark 
        \\
       
        & \cellcolor{Gray} MINS \cite{wang2022news}
        & \cellcolor{Gray} \Checkmark 
        & \cellcolor{Gray} \Checkmark 
        & \cellcolor{Gray} \Checkmark 
        & \cellcolor{Gray} \Checkmark 
        & \cellcolor{Gray} \Checkmark 
        & \cellcolor{Gray} \Checkmark 
        & \cellcolor{Gray} \Checkmark 
        \\
       
        & CenNewsRec \cite{qi2020privacy}
        & \Checkmark 
        & \Checkmark 
        & \Checkmark 
        & \Checkmark 
        & \Checkmark 
        & \Checkmark 
        & \Checkmark 
        \\

        \multirow{-10}{*}{\rotatebox[origin=c]{90}{\textbf{GeneralRec}}}
        & \cellcolor{Gray} MINER \cite{li2022miner}
        & \cellcolor{Gray} \Checkmark 
        & \cellcolor{Gray} \Checkmark 
        & \cellcolor{Gray} \Checkmark 
        & \cellcolor{Gray} \Checkmark 
        & \cellcolor{Gray} \Checkmark 
        & \cellcolor{Gray} \Checkmark 
        & \cellcolor{Gray} \Checkmark 
        \\
       
        \hdashline

        & SentiRec \cite{wu2020sentirec}
        & \Checkmark 
        & \Checkmark 
        & \Checkmark 
        & \Checkmark 
        & \Checkmark 
        & \Checkmark 
        & \Checkmark 
        \\
       
        & \cellcolor{Gray} SentiDebias \cite{wu2022removing}
        & \cellcolor{Gray} \Checkmark 
        & \cellcolor{Gray} \Checkmark 
        & \cellcolor{Gray} \Checkmark 
        & \cellcolor{Gray} \Checkmark 
        & \cellcolor{Gray} \Checkmark 
        & \cellcolor{Gray} \XSolidBrush 
        & \cellcolor{Gray} \XSolidBrush 
        \\

        \multirow{-3}{*}{\rotatebox[origin=c]{90}{\textbf{FairRec}}}
        & MANNeR (\cite{iana2023train})
        & \Checkmark 
        & \Checkmark 
        & \Checkmark 
        & \Checkmark 
        & \XSolidBrush 
        & \Checkmark 
        & \XSolidBrush 
        \\

        \hline
        \end{tabular}%
    }   
    \caption{List of currently available models in \newsreclib{}, and supported configurations. For \textit{click behavior fusion} we differentiate between \texttt{early fusion} (EF) and \texttt{late fusion} (LF). Models can be trained with \texttt{cross-entropy loss} (CE), \texttt{supervised contrastive loss} (SCL), and a \texttt{dual objective} combining both CE and SCL losses as weighted average (Dual). The dashed line separates the general (\texttt{GeneralRec}) from the fairness-aware  (\texttt{FairRec}) recommendation models.}
    \label{tab:model_configs}
\end{table*}

Additionally, \newsreclib{} integrates 3 fairness-aware models, namely NNRs that target diversity of recommendations along with pure content-based personalization:
\begin{itemize}
    \item \textit{SentiRec} \cite{wu2020sentirec} uses a similar architecture to NRMS \cite{wu2019nrms} and injects sentiment information by optimizing simultaneously for content personalization, as well as sentiment prediction. Additionally, it regularizes the NNR for sentiment diversity. 
    
    \item \textit{SentiDebias} \cite{wu2022removing} is a framework for sentiment debiasing which uses the architecture of NRMS \cite{wu2019nrms}, as well as adversarial learning to reduce the model's sentiment bias (originating from the user data) and generate sentiment-agnostic and diverse recommendations.
    
    \item \textit{MANNeR} \cite{iana2023train} is a modular framework for multi-aspect neural news recommendation, which comprises two types of modules, each with a corresponding NE (which combines a PLM-based text encoder with an entity embedder consisting of a pretrained embedding and multi-head self-attention layer), which are responsible for content-based, and respectively, aspect-based personalization. Both modules are trained with a contrastive metric objective. MANNeR uses late fusion \cite{iana2023simplifying} instead of standard user encoders. At inference time, the aspect-specific similarity scores are arbitrarily aggregated depending on the downstream task (e.g., content-based personalization, aspect-based diversification) to produce a final ranking of the news.
\end{itemize}

Table \ref{tab:model_configs} provides an overview of the supported configurations for the available models. For each model, users can choose the type of news encoder, click behavior fusion, and training objective. Note that for some models, due to the high interdependencies between NE and UE, it is not possible to easily replace the original NE with a PLM-based one without breaking the framework's modularity. Similarly, some models have been designed from the start with a PLM-based NE. In both of these cases, we only provide support for the original NE. Due to the design of some model architectures, changing the training objective would modify the functionality of the model (e.g., using different loss functions in the \texttt{CR-Module} and \texttt{A-Module} of MANNeR \cite{iana2023train}). In these cases, we only provide support for one training objective.

\end{document}